\documentclass{PoS}
\usepackage{epstopdf}
\usepackage[T1]{fontenc}
\usepackage{extarrows}
\usepackage{bm,amsmath,amssymb}
\usepackage[mathscr]{eucal}
\usepackage{graphicx}
\usepackage{cancel}

\newcommand{\nlo}{{\rm \scriptscriptstyle NLO}}

\newcommand{\beq}{\begin{equation}}
\newcommand{\eeq}{\end{equation}}

\newcommand{\order}[1]{\mathcal{O}{(#1)}}

\newcommand{\bx}{\bm{x}}
\newcommand{\by}{\bm{y}}

\newcommand{\bz}{\bm{z}}
\newcommand{\bw}{\bm{w}}

\title{Forward trijet production in proton-nucleus collisions}

\ShortTitle{}

\author{\speaker{Y. Mulian}\\
        Institut de Physique Theorique, Universite Paris Saclay, CNRS, CEA, F-91191, Gif-sur-Yvette, France\\
        E-mail: \email{yair25m@gmail.com}}

\abstract{Using the formalism of the light-cone wave function in perturbative QCD together with the hybrid factorization, we compute the cross-section for three particle production at forward rapidities in proton-nucleus collisions. In this picture, the three produced partons --- a quark accompanied by a gluon pair, or two quarks plus one antiquark --- are all generated via one or two successive splittings of a quark from the incoming proton, that was originally collinear with the latter. The three partons are put on-shell by their scattering off the nuclear target, described as Lorentz-contracted shockwave.
We explicitly compute the three-parton Fock space components of the light-cone wave function of the incoming quark and its outgoing state, which encodes the information both on the evolution in time as well as the scattering process. This outgoing state is also an ingredient for other interesting calculations, like the next-to-leading order correction to the cross-section for the production of a pair of jets.}

\FullConference{XXVI. International Workshop on Deep-Inelastic Scattering and Related Subjects,\\
		16 April - 20 April 2018\\
		Kobe, Japan}

\begin{document}
\section{Introduction to Gluon Saturation}
Particle production in proton-nucleus collisions at forward rapidities (in the proton fragmentation region) represents an important source of information about the small-$x$ part of the nuclear wavefunction, where gluon occupation numbers are high and non-linear effects like gluon saturation and multiple scattering are expected to be important. Within perturbative QCD, the corresponding cross-sections can be computed using the Colour Glass Condensate (CGC) effective theory \cite{Gelis:2010nm}, which is currently known to next-to-leading order (NLO) accuracy (at least for the high-energy evolution and for specific scattering processes), together with the so-called ``hybrid factorization''  \cite{Dumitru:2005gt}. The physical picture underlying this factorization is that the ``forward'' jets (or hadrons) observed in the final state are generally produced via the fragmentation of a single collinear parton from the incoming proton, which carries a large fraction $x_p\sim\order{1}$ of the longitudinal momentum of the proton and here is assumed to be a quark.

Using this approach, one has so far computed the cross-section for single inclusive hadron production, first  to leading-order (LO) accuracy \cite{Kovchegov:1998bi,Kovchegov:2001sc,Dumitru:2002qt} and then to NLO \cite{Chirilli:2011km,Chirilli:2012jd,Altinoluk:2014eka,Iancu:2016vyg}, and that for di-jet production only at LO \cite{Baier:2005dv,Marquet:2007vb,Dominguez:2011wm,Iancu:2013dta,vanHameren:2016ftb}. The results thus obtained compare quite well with the phenomenology, for both the single inclusive spectra \cite{Albacete:2003iq,Kharzeev:2003wz,Iancu:2004bx,Blaizot:2004wu,Blaizot:2004wv,Dumitru:2005gt,Albacete:2010bs,Tribedy:2011aa,Rezaeian:2012ye,Lappi:2013zma,Stasto:2013cha} and the di-jet production\cite{Albacete:2010pg}.

The aim of this proceeding is to explain how to compute multi-particle production cross sections in proton-nucleus collisions at forward rapidities, that is, in the fragmentation region of the proton projectile. Our dominant contribution comes from the process where a valence quark from the proton, possibly accompanied by its radiation products, scatters off the gluon distribution in the nucleus and then emerges in the final state.  We shall compute this process within perturbative QCD, so in particular we shall ignore confinement: our `final state' will be built with partons (quarks and gluons), rather than physical hadrons.
\section{The Outgoing State Formalism}

In order to be able to describe a scattering process we should understand both how the incoming state evolves with time, and how does it interact with the target. The state which encodes the information about the time evolution of an initial bare quark state $\left|q_{\lambda}^{\alpha}(q^{+},\,\bm{q})\right\rangle$ ($\alpha$ and $\lambda$ denote the color and polarization indices, while $q$ is its momenta) is given by:
\begin{equation}\label{loquar}
\left|q_{\lambda}^{\alpha}(q^{+},\,\bm{q})\right\rangle _{in}\,\equiv\,U(0,\,-\infty)\,\left|q_{\lambda}^{\alpha}(q^{+},\,\bm{q})\right\rangle 
\end{equation}
where $U$ is a unitary evolution operator, defied by:
\begin{equation}
U(t,\,t_{0})\,=\,\mathrm{T}\,\exp\left\{ -i\int_{t_{0}}^{t}dt_{1}\:H_{I}(t_{1})\right\} 
\end{equation}
At leading order, as the incoming bare quark WF evolves with time, it can emit a gluon (we treat the kinematics exactly, assuming no approximation for the emission vertices). The reader can find the result for the LO incoming bare quark WF as well as the cross section for the forward dijet production in \cite{Marquet:2007vb}. As a result of the collision, the partonic system also acquires a {\it total} transverse momentum of the order of the saturation momentum in the nucleus. In the high-energy regime of interest, the effects of multiple scattering can be resumed to all orders by using the eikonal approximation (a parton from the projectile does not get deflected, but merely acquires a color rotation.) This amounts to associating a Wilson line \cite{Lublinsky:2016meo} built with the colour field of the target to each parton from the projectile (the operator which assigns the Wilson lines for each parton is denoted here by $\hat{S}$). While at leading order the procedure to insert the shockwave is straightforward, it cannot be easily generalized for higher orders. An elegant and systematical way to generate at all the different contributions from the possible locations in which the interaction may occur is given by the expression for the outgoing state:
\begin{equation}\begin{split}\label{scatquar}
&\left|q_{\lambda}^{\alpha}\right\rangle _{out}\,\equiv\,U(\infty,\,0)\,\hat{S}\,U(0,\,-\infty)\,\left|q_{\lambda}^{\alpha}\right\rangle .
\end{split}\end{equation}
The last expression can be computed perturbativly:
\begin{equation}
\left|q_{\lambda}^{\alpha}\right\rangle _{out}\,=\,\left|q_{\lambda}^{\alpha}\right\rangle \,+\,\left|q_{\lambda}^{\alpha}\right\rangle _{out}^{(g)}\,+\,\left|q_{\lambda}^{\alpha}\right\rangle _{out}^{(g^{2})}\,+\,\ldots\,,
\end{equation}
Since here we are looking for the situation in which we have at least three partons at the final state, we have to compute the outgoing state up to order $g^{2}$:
\begin{equation}\begin{split}\label{out1}
\left|q_{\lambda}^{\alpha}\right\rangle _{out}^{(g)}\,=\,-\sum_{i\neq f}\left|f\right\rangle \left\langle f\left|S\right|i\right\rangle \frac{\left\langle i\left|H_{{\rm int}}\right|in\right\rangle }{E_{i}-E_{in}}\,+\,\sum_{i\neq f}\left|f\right\rangle \frac{\left\langle f\left|H_{{\rm int}}\right|i\right\rangle }{E_{f}-E_{i}}\left\langle i\left|S\right|in\right\rangle ,
\end{split}\end{equation}
\begin{eqnarray}\label{outgoing}
&&\left|q_{\lambda}^{\alpha}\right\rangle _{out}^{(g^{2})}\,=\,\sum_{i\neq in,j\neq in}\left|f\right\rangle \left\langle f\left|S\right|j\right\rangle \frac{\left\langle j\left|H_{{\rm int}}\right|i\right\rangle \,\left\langle i\left|H_{{\rm int}}\right|in\right\rangle }{(E_{j}-E_{in})(E_{i}-E_{in})}\,+\,\sum_{i\neq j,j\neq f}\left|f\right\rangle \frac{\left\langle f\left|H_{{\rm int}}\right|j\right\rangle \,\left\langle j\left|H_{{\rm int}}\right|i\right\rangle }{(E_{f}-E_{j})(E_{f}-E_{i})}\,\left\langle i\left|S\right|in\right\rangle \nonumber\\
&&-\sum_{i\neq in,j\neq f}\left|f\right\rangle \frac{\left\langle f\left|H_{{\rm int}}\right|j\right\rangle }{E_{f}-E_{j}}\left\langle j\left|S\right|i\right\rangle \frac{\left\langle i\left|H_{{\rm int}}\right|in\right\rangle }{E_{i}-E_{in}}.
\end{eqnarray}
where $H_{{\rm int}}$ denotes the interaction part of the QCD Hamiltonian in light-cone gauge, and we should sum over all the possible states $\left|i\right\rangle$, $\left|j\right\rangle$, and $\left|f\right\rangle$, of our Fock space (the Fock space here consists of the quark state, quark and a gluon state, quark and two gluons state, and two quarks and an anti-quark). The respective energies for the states mentioned are denoted by $E_{i}, E_{j}, E_{f}$, and the state $\left|in\right\rangle $ denotes the incoming state (which in our case is a bare quark state). After summing over the different states, it can be seen that the state in eq. (\ref{outgoing}) has the following structure:
\begin{equation}\begin{split}\label{prodlo}
&\left|q_{\lambda}^{\alpha}\right\rangle _{out}^{(g^{2})}\,\simeq\,\hat{Z}_{\nlo}\left|q_{\lambda}^{\alpha}\right\rangle \,+\,\left|\psi_{\lambda}^{\alpha}\right\rangle _{qg}\,+\,\left|\psi_{\lambda}^{\alpha}\right\rangle _{qq\overline{q}}\,+\,\left|\psi_{\lambda}^{\alpha}\right\rangle _{qgg}.
\end{split}\end{equation}
Where $\hat{Z}_{\nlo}$ accounts for the normalization of the WF and the partons produced at the final state appear as a subscript.
\subsection{Computing the Outgoing State}
\begin{figure}[!t]
\includegraphics[scale=0.58]{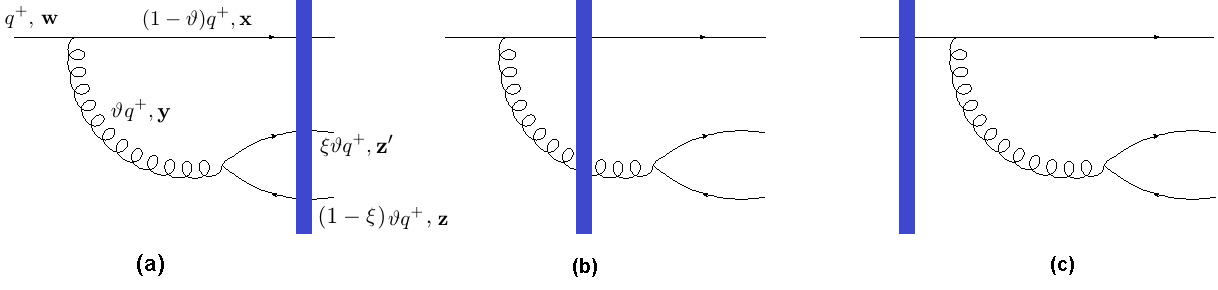}
\caption{The three possible configurations for the interplay between the quark evolution and the scattering, for a final partonic state built with three quarks and a propagating intermediate gluon: (a) initial-state evolution, (b) mixed (the gluon emission occurs prior to scattering, but its splitting happens after the scattering), (c) final-state evolution.}
\label{quarinsert}
\end{figure}

As mentioned in the introduction, out interest is to compute the leading-order cross-section for producing three partons in the final state. In order to demonstrate the method, it is enough to focus on the case in which two quarks and an anti-quark are produced at the final state (along with that contribution we can also have one quark together with two additional gluons that will not be discussed here). To lowest order in perturbation theory, the incoming state built with these 3 (bare) partons involves either one or two emission vertices, which we denote here as regular and instantaneous emission (in the instantaneous channel an intermediate gluon is not created, and the quark anti-quark pair are emitted directly from the incoming state.) The total contribution is a sum of the two contributions, $\left|\psi_{\lambda}^{\alpha}\right\rangle _{qq\overline{q}}\,\equiv\,\left|\psi_{\lambda}^{\alpha}\right\rangle _{qq\overline{q}}^{inst}\,+\,\left|\psi_{\lambda}^{\alpha}\right\rangle _{qq\overline{q}}^{reg}$. In what follows we shall deal only with the contribution from the regular emission. The contribution from this channel to the outgoing state in eq. (\ref{outgoing}) is given by the following expression \cite{paper}:

\begin{eqnarray} \label{fin.qqq1}
&&\left|\psi_{\lambda}^{\alpha}(q^{+},\,\bm{w})\right\rangle _{qq\bar{q}}^{reg}\,=\,-\,\int_{\bm{x},\,\bm{z},\,\bm{z}^{\prime}}\,\int_{0}^{1}d\vartheta\,d\xi\:\frac{g^{2}\,\varphi_{\lambda_{2}\lambda_{3}}^{il}(\xi)\,\phi_{\lambda_{1}\lambda}^{ij}(\vartheta)\,\bm{Z}^{l}\,\left(\bm{X}^{j}+\xi\bm{Z}^{j}\right)\,q^{+}}{16\pi^{3}\left(\bm{X}+\xi\bm{Z}\right)^{2}\,\bm{Z}^{2}}\nonumber\\
&&\times\left[\Theta_{1}\,V^{\varrho\delta}(\bm{z}^{\prime})\,t_{\delta\epsilon}^{a}\,V^{\dagger\epsilon\rho}(\bm{z})\,V^{\sigma\beta}(\bm{x})\,t_{\beta\alpha}^{a}\,+\,\Theta_{2}\,t_{\varrho\rho}^{a}\,t_{\sigma\beta}^{a}\,V^{\beta\alpha}(\bm{w})-\,t_{\varrho\rho}^{b}\,V^{\sigma\beta}(\bm{x})\,U^{ba}(\bm{y})\,t_{\beta\alpha}^{a}\right]\nonumber\\
&&\times\delta^{(2)}\left(\bm{w}-\bm{C}\right)\left|\bar{q}_{\lambda_{3}}^{\rho}((1-\xi)\vartheta q^{+},\,\bm{z})\,q_{\lambda_{2}}^{\varrho}(\xi\vartheta q^{+},\,\bm{z}^{\prime})\,q_{\lambda_{1}}^{\sigma}((1-\vartheta)q^{+},\bm{x})\right\rangle ,
 \end{eqnarray}
 where  $\bx$, and $\bz'$ denote the transverse coordinates of two final quarks, while $\bz$ is the transverse coordinate of the anti-quark. The transverse position of the intermediate gluon $\by$, and the corresponding position $\bw$ of the incoming quarks are given by
\begin{equation}\label{defyc}
\bm{y}\,\equiv\,\xi\bm{z}^{\prime}+(1-\xi)\bm{z}\,;\qquad\bm{w}\,=\,(1-\vartheta)\bm{x}\,+\,\xi\vartheta\bm{z}^{\prime}\,+\,(1-\xi)\vartheta\bm{z}.
\end{equation}
For compactness we also define:
\begin{equation}
\bm{X}\equiv\bm{x}-\bm{z};\qquad\bm{Z}\,\equiv\,\bm{z}\,-\,\bm{z}^{\prime},
\end{equation}
$U^{ab}(\bm{x})$ and $V^{\alpha\beta}(\bm{x})$ are Wilson lines in the adjoint and fundamental representations, and
\begin{equation}
\Theta_{1}\,\equiv\,\frac{(1-\vartheta)\left(\bm{X}+\xi\bm{Z}\right)^{2}}{(1-\vartheta)\left(\bm{X}+\xi\bm{Z}\right)^{2}\,+\,\xi(1-\xi)\bm{Z}^{2}};\qquad\Theta_{2}\,\equiv\,\frac{\xi(1-\xi)\bm{Z}^{2}}{(1-\vartheta)\left(\bm{X}+\xi\bm{Z}\right)^{2}\,+\,\xi(1-\xi)\bm{Z}^{2}},
\end{equation}
are fractions between $0$ and $1$. Note that the three terms inside the square brackets of Eq.~(\ref{fin.qqq1}) have a different color structure corresponding to the three possibilities for the insertion of the shockwave, as depicted on Fig.~\ref{quarinsert}.

\section{The Trijet Cross Section\label{trijetfinal}}
 The cross-section for partons--nucleus scattering is obtained by averaging number density operators (defined below) over all the colour field configurations in the target with the CGC weight function \cite{Gelis:2010nm}. In order to pass from the partonic cross section to a cross section which involves hadrons, it must be convoluted with the quark distribution function of the proton and the fragmentation functions for partons fragmenting into hadrons, or jets.

Within the hybrid factorization \cite{Dumitru:2005gt}, the cross-section for producing three jets at forward rapidities in proton-nucleus collisions and to leading order in pQCD is simply obtained by convoluting the respective partonic cross-section with the proton parton distribution functions for the partons which have initiated the process. 
\begin{equation}\label{trijets}
\frac{d\sigma^{pA\to3jet+X}}{d^{3}q_{1}\,d^{3}q_{2}\,d^{3}q_{3}}\,=\,\int dx_{p}\,q(x_{p},\mu^{2})\left(\frac{d\sigma^{qA\rightarrow qgg+X}}{d^{3}q_{1}\,d^{3}q_{2}\,d^{3}q_{3}}\,+\,\frac{d\sigma^{qA\rightarrow qq\overline{q}+X}}{d^{3}q_{1}\,d^{3}q_{2}\,d^{3}q_{3}}\right).
 \end{equation}
Here, $q_{1}$, $q_{2}$, $q_{3}$ are the momenta of the measured partons. $q(x_p,\mu^2)$ is the quark distribution function of the proton evaluated for a longitudinal momentum fraction $x_p=q^+/Q^+$ (with $Q^+$ the proton longitudinal momentum) and for a transverse (or virtuality) scale $\mu^2$. The value of $x_p$ is actually fixed by the $\delta$-function implicit in the partonic cross-sections and which expresses the conservation of longitudinal momentum  ($q^{+}=q_{1}^{+}+q_{2}^{+}+q_{3}^{+}$).

\begin{figure}[!h]\center
    \includegraphics[scale=0.6]{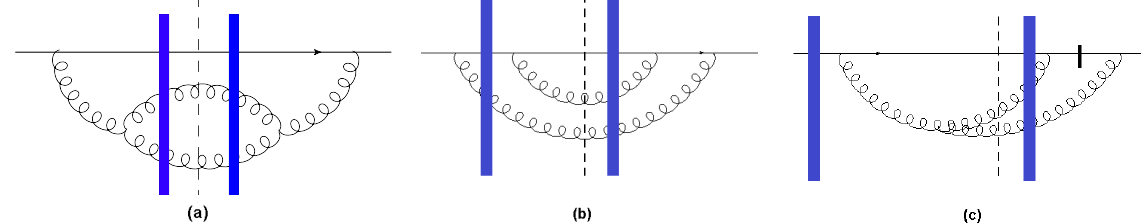}
  \caption{Three examples of diagrams which demonstrate the production of two gluons. The two direct contributions are shown in $(a)$ and $(b)$, in which the two gluons can be produced by the splitting of the original gluon or by emitting a second one from the incoming quark. Contribution $(c)$ represents the interference between these contributions.\label{qgg1fig1}}
\end{figure}

The three-parton cross-sections in eq. (\ref{trijets}) are in turn computed as expectation values over the outgoing--state of the product of three number-density Fock space operators for {\it bare} partons:
 \begin{equation}\begin{split}\label{qqqcross}
&\frac{d\sigma^{qA\rightarrow qq\overline{q}+X}}{d^{3}q_{1}\,d^{3}q_{2}\,d^{3}q_{3}}\,\equiv\,\frac{1}{2N_{c}\,L}\,_{qq\overline{q}}\left\langle \psi_{\lambda}^{\alpha}(q^{+},\,\bm{q}=0_{\perp})\right|\,\hat{\mathcal{N}}_{q}(q_{1})\,\hat{\mathcal{N}}_{q}(q_{2})\,\hat{\mathcal{N}}_{\overline{q}}(q_{3})\,\left|\psi_{\lambda}^{\alpha}(q^{+},\,\bm{q}=0_{\perp})\right\rangle _{qq\overline{q}}.\\
&=\,\frac{1}{2N_{c}\,L}\,\int_{\bm{w},\,\overline{\bm{w}}}\,{}_{qq\bar{q}}\left\langle \psi_{\lambda}^{\alpha}(q^{+},\,\overline{\bm{w}})\right|\,\hat{\mathcal{N}}_{q}(q_{1})\,\hat{\mathcal{N}}_{q}(q_{2})\,\hat{\mathcal{N}}_{\overline{q}}(q_{3})\,\left|\psi_{\lambda}^{\alpha}(q^{+},\,\bm{w})\right\rangle _{qq\bar{q}},
 \end{split}\end{equation}
where the number density operators for (bare) quarks, anti-quarks, and gluons are given by
\begin{equation}
\label{Nq}
\hat{\mathcal{N}}_{q}(p)\,\equiv\,\frac{1}{(2\pi)^{3}}\,b_{\lambda}^{\alpha\dagger}(p)\,b_{\lambda}^{\alpha}(p),\qquad\qquad\hat{\mathcal{N}}_{g}(k)\,\equiv\,\frac{1}{(2\pi)^{3}}\,a_{i}^{a\dagger}(k)\,a_{i}^{a}(k).
\end{equation}
It should be mentioned that the factor ${1}/{2N_{c}}$ in eq. (\ref{qqqcross}) accounts for the average over the colors and polarizations of the initial quark. The factor $1/L$, with $L$ denoting the {\em a priory} infinite extension of the longitudinal axis, is needed to remove an ill-defined delta function expression the conservation of the longitudinal momentum.
The gluons contribution to the trijet cross section $\frac{d\sigma^{qA\rightarrow qgg+X}}{d^{3}q_{1}\,d^{3}q_{2}\,d^{3}q_{3}}$ is given similarily by replacing the quark and an anti-quark number density operators by the corresponding gluonic ones (see figure \ref{qgg1fig1}).
\begin{figure}[!h]
    \includegraphics[scale=0.9]{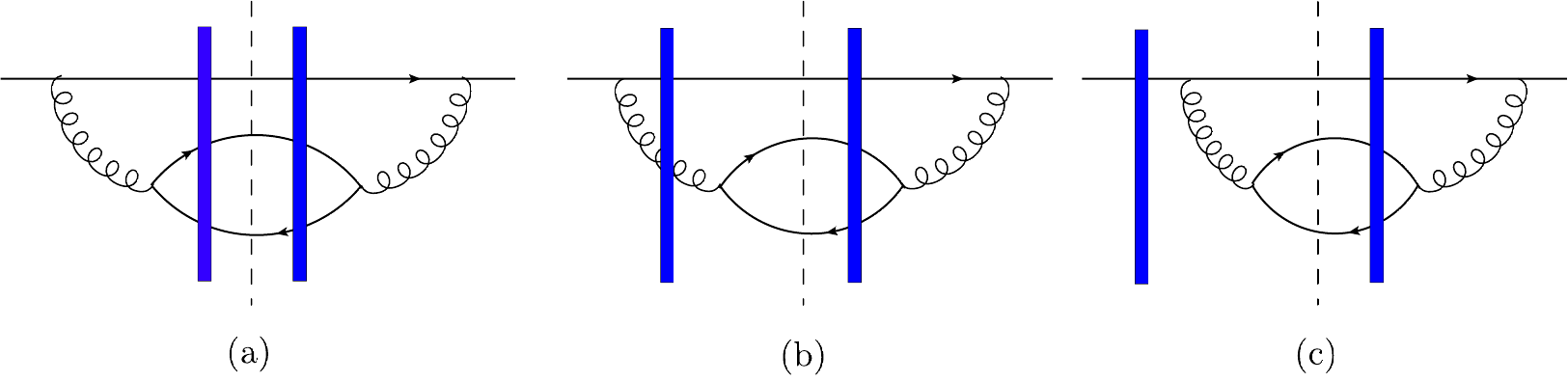}
  \caption{Three examples of diagrams which demonstrate the production of a quark antiquark pair via an intermediate gluon in the direct and conjugate amplitudes. In total there are 9 such contributions.\label{qqqfig1}}
\end{figure}

The contribution of the channel $qA\rightarrow qq\overline{q}+X$ to the trijet cross section, as given by eq. (\ref{qqqcross}), consists of four different parts. One of these parts involve the creation of a gluon in the direct and conjugate amplitudes before and after the splitting to quark and anti-quark pair (denoted by "reg-reg"). Another part involves the instantaneous creation of the quark anti-quark directly from the incoming quark (denoted by "inst-inst"). The two remaining contributions correspond to the interference between the regular and instantaneous emissions (denoted by "reg-inst" and "inst-reg"). Therefore, we can write the result in the following way:
 \begin{equation}\begin{split}
&\frac{d\sigma^{qA\rightarrow qq\overline{q}+X}}{d^{3}q_{1}\,d^{3}q_{2}\,d^{3}q_{3}}=\left.\frac{d\sigma^{qA\rightarrow qq\overline{q}+X}}{d^{3}q_{1}\,d^{3}q_{2}\,d^{3}q_{3}}\right|_{reg-reg}+2\mathrm{Re}\left.\frac{d\sigma^{qA\rightarrow qq\overline{q}+X}}{d^{3}q_{1}\,d^{3}q_{2}\,d^{3}q_{3}}\right|_{reg-inst}+\left.\frac{d\sigma^{qA\rightarrow qq\overline{q}+X}}{d^{3}q_{1}\,d^{3}q_{2}\,d^{3}q_{3}}\right|_{inst-inst}.
\end{split}\end{equation}
In order to express the result for the first term in the last equation, one has to  introduce two basic gauge-invariant operators, known as dipole and baryon: 
\begin{equation}\label{lowils2}
\mathcal{S}\left(\overline{\bm{w}},\,\bm{w}\right)\,\equiv\,\frac{1}{N_{c}}\,\mathrm{tr}\left[V^{\dagger}(\overline{\bm{w}})\,V(\bm{w})\right],\qquad\mathcal{Q}\,(\overline{\bm{x}},\,\bm{x},\,\bm{z},\,\overline{\bm{z}})\,\equiv\,\frac{1}{N_{c}}\,\mathrm{tr}\left[V^{\dagger}(\overline{\bm{x}})\,V(\bm{x})\,V^{\dagger}(\bm{z})\,V(\overline{\bm{z}})\right].
\end{equation}

After inserting the result in eq. (\ref{outgoing}) to the definition of the cross section (\ref{qqqcross}), and retaining only the large $N_{c}$ limit contributions, the result can be expressed solely in terms of the dipole and quadropole:
\begin{eqnarray}
&&\left.\frac{d\sigma^{qA\rightarrow qq\overline{q}+X}}{d^{3}q_{1}\,d^{3}q_{2}\,d^{3}q_{3}}\right|_{reg-reg}\\
&&\equiv\,\frac{\alpha_{s}^{2}\,N_{c}\,N_{f}}{256\pi^{5}(q^{+})^{2}}\,\delta(q^{+}-q_{1}^{+}-q_{2}^{+}-q_{3}^{+})\,\int_{\bm{\overline{x}},\,\bm{\overline{z}},\,\bm{\overline{z}}^{\prime},\,\bm{x},\,\bm{z},\,\bm{z}^{\prime}}\,e^{-i\bm{q}_{1}\cdot(\bm{x}-\bm{\overline{x}})-i\bm{q}_{2}\cdot(\bm{z}-\bm{\bar{z}})-i\bm{q}_{3}\cdot(\bm{z}^{\prime}-\bm{\bar{z}}^{\prime})}\nonumber\\
&&\times K_{qq\overline{q}}\left(\bm{\overline{x}},\,\bm{\overline{z}},\,\bm{\overline{z}}^{\prime},\,\bm{x},\,\bm{z},\,\bm{z}^{\prime}\right)\left[\overline{\Theta}_{1}\,\Theta_{1}\,\mathcal{Q}(\overline{\bm{x}},\,\bm{x},\,\bm{z}^{\prime},\,\overline{\bm{z}}^{\prime})\,\mathcal{S}(\overline{\bm{z}},\,\bm{z})\,-\,\overline{\Theta}_{1}\,\mathcal{Q}(\overline{\bm{x}},\,\bm{x},\,\bm{y},\,\overline{\bm{z}}^{\prime})\,\mathcal{S}(\overline{\bm{z}},\,\bm{y})\right.\nonumber\\
&&-\,\Theta_{1}\,\mathcal{Q}(\overline{\bm{x}},\,\bm{x},\,\bm{z}^{\prime},\,\overline{\bm{y}})\,\mathcal{S}(\overline{\bm{y}},\,\bm{z})\,+\,\overline{\Theta}_{2}\,\Theta_{1}\,\mathcal{S}(\overline{\bm{w}},\,\bm{z})\,\mathcal{S}(\bm{z}^{\prime},\,\bm{x})\,+\,\overline{\Theta}_{1}\,\Theta_{2}\,\mathcal{S}(\overline{\bm{x}},\,\bm{\overline{z}}^{\prime})\,\mathcal{S}(\bm{\overline{z}},\,\bm{w})\nonumber\\
&&\left.\,+\,\mathcal{Q}(\overline{\bm{x}},\,\bm{x},\,\bm{y},\,\overline{\bm{y}})\,\mathcal{S}(\overline{\bm{y}},\,\bm{y})-\,\overline{\Theta}_{2}\,\mathcal{S}(\overline{\bm{w}},\,\bm{x})\,\mathcal{S}(\bm{x},\,\bm{y})\,-\,\Theta_{2}\,\mathcal{S}(\overline{\bm{x}},\,\bm{\overline{y}})\,\mathcal{S}(\bm{\overline{y}},\,\bm{w})\,+\,\overline{\Theta}_{2}\,\Theta_{2}\mathcal{S}\left(\bm{\overline{w}},\,\bm{w}\right)\right]\nonumber\\
&&\,+\,\left(q_{1}^{+}\leftrightarrow q_{2}^{+},\:\bm{q}_{1}\leftrightarrow\bm{q}_{2}\right).\nonumber
\end{eqnarray}

\providecommand{\href}[2]{#2}\begingroup\raggedright\endgroup

\end{document}